\documentclass[conference]{IEEEtran}
\pdfoutput=1

\usepackage{cite}
\usepackage{amsmath,amssymb,amsfonts}
\usepackage{mathtools}
\usepackage{graphicx}
\usepackage{textcomp}
\usepackage{xcolor}
\usepackage{algorithm}
\usepackage{algpseudocode}
\usepackage{hyperref}
\hypersetup{
  colorlinks   = true,    % Colours links instead of ugly boxes
  urlcolor     = black,    % Colour for external hyperlinks
  linkcolor    = black,    % Colour of internal links
  citecolor    = blue      % Colour of citations
}

\def\BibTeX{{\rm B\kern-.05em{\sc i\kern-.025em b}\kern-.08em
    T\kern-.1667em\lower.7ex\hbox{E}\kern-.125emX}}

\DeclareMathOperator*{\argmax}{arg\,max}

\begin{document}
\title{Dynamic Collaborative Filtering for Matrix- and Tensor-based Recommender Systems%*\\
}

%\author{
%\IEEEauthorblockN{Anonymous}
%\IEEEauthorblockA{Anonymous} \\
%}
%\and
\author{
\IEEEauthorblockN{Albert Saiapin}
\IEEEauthorblockA{ 
Delft University of Technology\\
Delft, The Netherlands \\
a.saiapin@tudelft.nl}
\and
\IEEEauthorblockN{Ivan Oseledets}
\IEEEauthorblockA{
Skolkovo Institute of Science and \\
Technology\\
Moscow, Russia \\
i.oseledets@skoltech.ru}
\and
\IEEEauthorblockN{Evgeny Frolov}
\IEEEauthorblockA{
Skolkovo Institute of Science and \\
Technology\\
Moscow, Russia \\
evgeny.frolov@skoltech.ru}
}
%\and
%\and
%\and

\maketitle

\begin{abstract}
In production applications of recommender systems, a continuous data flow is employed to update models in real-time. Many recommender models often require complete retraining to adapt to new data. In this work, we introduce a novel collaborative filtering model for sequential problems known as Tucker Integrator Recommender - TIRecA. TIRecA efficiently updates its parameters using only the new data segment, allowing incremental addition of new users and items to the recommender system. To demonstrate the effectiveness of the proposed model, we conducted experiments on four publicly available datasets: MovieLens 20M, Amazon Beauty, Amazon Toys and Games, and Steam. Our comparison with general matrix and tensor-based baselines in terms of prediction quality and computational time reveals that TIRecA achieves comparable quality to the baseline methods, while being 10-20 times faster in training time.
\end{abstract}


\section{Introduction}
Recommender systems have become integral technologies across a multitude of applications, significantly enhancing user experiences and delivering substantial business value, as exemplified by industry leaders such as Amazon and YouTube.

Yet, a fundamental challenge persists in this dynamic landscape: How can we effectively refine model parameters within the continuous flow of data? Envision the life cycle of a production-level recommender system—a process involving two critical stages: using the latest model for predicting user preferences and the rigorous task of retraining or updating the model to adapt to the ever-evolving data flow.

At the core of this challenge is the optimization of model retraining. It's a quest to adjust model parameters seamlessly, aligning with the data dynamics, all while preserving the patterns within the structure of model weights. This is a routine procedure every recommender model performs to be up to date. However, the latest state-of-the-art models~\cite{McAuley_SASRec_2018, DuoRec, Eff_SASREC} cannot deal with the problem of adding new users or items to a recommender system because the whole neural network architecture changes.

In this paper, we choose a different path. We initiate our exploration by addressing the problem formulation of the PureSVD model~\cite{Cremonesi_Topn_2010} and progressively refine it to enable the effective updating of user and item embeddings in response to incoming data.

% RELATIONSHIP OF THIS STUDY TO PREVOUSLY PUBLISHED
The main idea we start from is to incorporate the notion of time into the optimization problem formulation and then use the same techniques as the authors of~\cite{Olaleke_Dynamic_Modeling_2021} applied. However, we extend their model by processing new users and items, and generating embeddings for them in an incremental fashion.

In addition to it, as was mentioned above, sequential models based on Neural Networks show state of the art quality on the task of the next item prediction~\cite{McAuley_SASRec_2018, S3Rec, DuoRec}. However, as the authors of~\cite{frolov_tensorbased_2022} showed, properly prepared factorization models can show the same or superior quality.
With this in mind, we propose a model that can use sequential information about user preferences based on GA-SATF~\cite{frolov_tensorbased_2022} model. It requires a more complex model formulation compared PureSVD~\cite{Cremonesi_Topn_2010}. Namely, we are using a three-dimensional tensor data representation as there are three different modes in the data: a user, an item, and a sequential order. 

% TASKS TO ACHIEVE THE GOAL
The main contributions~\footnote{\url{https://github.com/AlbMLpy/DynamicCF}} of this work can be summarized as
follows:
\begin{enumerate}
    \item We develop a matrix-based dynamic collaborative filtering recommender system using Projector Splitting Integrator (PSI). The model can process new users and items on the fly.
    \item We develop a novel tensor-based sequential collaborative filtering recommender system using Tucker Integrator (TI). The model can process new users and items dynamically.
    \item We suggest and experimentally check several methodologies to initialize and further update embeddings (model factors) for new users and new items.
\end{enumerate}

% PRACTICAL VALUE OF THE RESEARCH
As a result, the developed models could potentially replace many models in production, from PureSVD~\cite{Cremonesi_Topn_2010} to SASRec~\cite{McAuley_SASRec_2018}. They could be faster than state-of-the-art neural networks, save money, reduce time consumption, and preserve the same or comparable accuracy. On the other hand, these models could emphasize the importance of developing general-purpose mathematical frameworks and continuing the research in the dynamical modeling and tensor factorization areas as an alternative to mainstream research directions based on neural networks.

\section{Related Work}
\subsection{Recommender Systems}\label{sec:recsys}
There are different models in recommender systems field. Some of them rely on content information \cite{Pazzani_Content_based_RS_2007} while others on collaborative signals \cite{Schafer_CF_2007}. In our study we will focus on collaborative filtering for implicit data that can be sequential \cite{Donkers_Seq_RNN_2017}. 
Moreover, the authors of \cite{Cremonesi_Topn_2010} suggest to use top-n problem formulation where the goal of the recommender system is to find a few
specific items which are supposed to be most relevant to
the user. As this problem formulation better reflects commercial applications we would rely on it in this work.

\subsection{Low-Rank Approximations}\label{sec:low_rank_appr}
This section reviews the research related to
matrix and tensor decompositions which are the basis of this work.
The authors of \cite{Kolda_TD_2009} give a
comprehensive overview of general tensor decompositions and mathematical concepts behind them. Nevertheless, in order to handle big data and compute tensor decomposition out of it one should use different randomization techniques. One can find an extensive review of recent advances in randomization for the computation of a Tucker decomposition and Higher Order SVD in \cite{Phan_randomized_TD_2021}.
The connection between recommender systems field and tensor decompositions can be found in \cite{Frolov_TMRS_2016}.

However, in order to solve the dynamical problem formulation, which will be presented later, one have to explore the notion of dynamical low-rank approximations. Several studies \cite{Koch_DLRA_2007, Lubich_PSI_2013} present an increment-based computational approach for the low-rank approximation of time-dependent data matrices and solutions to matrix differential equations.
It is shown that the approach gives locally quasi-optimal low-rank approximations. A recent study~\cite{Olaleke_Dynamic_Modeling_2021} concluded that it is possible to integrate dynamical matrix factorization framework into a recommender system in order to get more stable solution.

The following studies~\cite{Koch_DTD_2010, Lubich_2015_MCTDHM} present high-dimensional problem formulation. The authors try to solve the same problem of the approximation of time-dependent data tensors and of solutions to tensor differential equations by tensors of low Tucker rank. This approach works with the increments
rather than the full tensor. In this
method, the derivative is projected onto the tangent space of the manifold of tensors represented in Tucker format.

Another area of low-rank approximations research is connected with incremental parameters updates. For example, the question of how an SVD may be updated by adding new rows and/or columns of data,
which may be missing values or contaminated with correlated noise is considered in \cite{brand_ISVD_2002, Nikolakopoulos_proj_svd_2021}. The size of the data matrix need not be known as an SVD is developed as the new data comes in. The paper~\cite{Hu_ITD_2011}, describes an online tensor subspace learning algorithm which models appearance changes by incrementally learning a tensor subspace representation through adaptively updating the sample mean and an eigenbasis for each unfolding matrix of the tensor.

\subsection{Sequential Models}\label{sec:seq_models}
Another substantial part of this research is connected with sequential recommender systems. For example, in \cite{McAuley_SASRec_2018} the authors proposed a self-attention based sequential model (SASRec) that allows to capture long-term semantics (like an RNN), but, using an attention mechanism, makes its predictions depend on relatively few actions (like a
Markov Chain). At each time step, SASRec seeks to identify which items are ‘relevant’ from a user’s action history, and use them to predict the next item. 
Top-n sequential recommendation models represent each user as a sequence of items he interacted in the past and aims to predict top-n ranked items that a user will likely interact in a “near future”. The order of interaction implies that sequential patterns play an important
role where more recent items in a sequence have a larger impact on the next item. 

In order to improve sequential recommender system performance the authors of~\cite{S3Rec} suggest to use the internal correlations in data to improve the data representations via pre-training methods. They consider several self-supervised objectives
to learn the correlations among attribute, item, subsequence, and sequence by using the mutual information maximization (MIM)
principle which is particularly suitable
in their scenario. The experimental results show better prediction quality results for S3Rec than for all the baselines including SASRec.
One more example of a neural sequential recommender could be - DuoRec~\cite{DuoRec}. This model uses a contrastive objective to address the representation degeneration problem by  putting regularization over sequence representations. This model shows superior performance to SASRec and S3Rec models.

Previous studies have almost exclusively focused on neural models to model sequential behaviour of a user with competitive prediction quality.
A recent study~\cite{frolov_tensorbased_2022} develops the model that mimics transformer self-attention with an alternative and more lightweight approach. It is a tensor factorization-based model that ingrains the structural knowledge about sequential data within the learning process. The authors demonstrate how certain properties of a self-attention network can be reproduced using special Hankel matrix representation. The resulting model has a shallow linear architecture and compares competitively to its neural counterpart.

\subsection{Literature Review Conclusions}\label{sec:lr_conclusion}
In this work we concentrate on the top-n problem formulation for sequential data with implicit feedback. It is more production oriented setting that can reflect the real performance of the recommender system.
The main model we develop will be based on several distinct components:
\begin{itemize}
    \item Tensor model for sequential recommendations - GA-SATF - based on \cite{frolov_tensorbased_2022}.
    \item Computational core of matrix and tensor decompositions - based on \cite{Phan_randomized_TD_2021}.
    \item Dynamical modelling integrators - based on \cite{Lubich_PSI_2013, Lubich_TI_2017}
    \item Addition of new users and items - based on \cite{brand_ISVD_2002}
\end{itemize}

\section{Problem Statement}\label{sec:prob_form}
The primary objective of a general recommender system is to leverage prior information about users, items, and additional contextual knowledge to accurately predict the most relevant items for a selected user. The relevance of items is typically measured using a relevance score function denoted as $f_{R}$, which is estimated from implicit feedback provided by users.

This study is an attempt to address the issue of sequence-aware tensor factorization for the next item prediction. The data utilized in this research can be organized in the form of a third order tensor $\mathcal{X}$ such that:
\begin{equation}\label{eq:data_tensor}
    x_{i j k} = \begin{cases}
       1, & \text{if j-th item is in position $p^i_{k}$ of i-th user}\\
        0,  & \text{otherwise},
    \end{cases}
\end{equation}
where $k = p_k^i - n_i + L$ and $n_i$ is the total number of
items in the ordered history of user i. To ensure consistency, any user history with a length exceeding $L$ is truncated, such that $n_i \le L$ for all users. Note that, regardless of the length of the original sequence, the most recent item is always located at position $L$. If a sequence contains fewer that $L$ items, zero padding is employed to fill in the remaining positions up to $L$.

The aforementioned data representation allows to define a tensor model using relevance score function: 
\begin{equation}
    \label{eq:rel_score_tensor}
    f_R: \text{User} \times \text{Item} \times \text{Position} \rightarrow \text{Relevance}
\end{equation}
that assigns some relevance score to each triplet of the
observed interactions between users and items considering their respective positions in the user's history of interactions.
As user consumption behavior may exhibit variability and the history length could potentially approach the size of the entire item catalog, computational challenges may arise due to the high dimensionality of positional encoding. However, it is reasonable to suppose that only a relatively small number of the most recent items significantly contribute to explaining current user decisions. Hence, truncating user sequences to a length of $L$, which is considerably smaller than the catalog size $N$, is a pragmatic approach.

In order to represent the tensor model \eqref{eq:rel_score_tensor} we will use the mathematics of Low-Rank Tensor Decomposition~\cite{Kolda_TD_2009}.
Apart from that, we suppose that the incoming tensor data depends on time. 
That is why, to encode the dynamical nature of the system we will use the theory from \cite{Lubich_TI_2017, Koch_DTD_2010}.
We would consider the approximation of time-varying tensors by tensors of low Tucker rank. The approach to get tensor decomposition factors is a continuous-time updating technique, which works only with the increments in the tensors rather than the tensors themselves and does not require computing any decompositions of large tensors. 

Consider a time-varying family of tensors $\mathcal{X}(t) \in \mathbb{R}^{M \times N \times L}$ for $0 \leq t \leq \hat{t}$.
Let $\mathcal{M}_r$ denote the manifold of all order-3 tensors of Tucker rank \textbf{r} = $(r_1, r_2, r_3)$, where $r_1 \leq M$, $r_2 \leq N$, $r_3 \leq L$. The best approximation to $\mathcal{X}(t)$ in $\mathcal{M}_r$ with respect to Frobenius norm is:
\begin{equation} 
\|\mathcal{X}(t) - \mathcal{R}(t)\| \rightarrow
\min_{\mathcal{R}(t) \in \mathcal{M}_r}
\end{equation}
where $\mathcal{R}(t) = \mathcal{C}(t)\times_{1}U(t)\times_{2}V(t)\times_{3}W(t)$ - Tucker decomposition~\cite{Kolda_TD_2009}, the factor matrices $U(t) \in \mathbb{R}^{M\times r_1}$, $V(t) \in \mathbb{R}^{N\times r_2}$, $W(t) \in \mathbb{R}^{L\times r_3}$ represent user, item, positional embeddings accordingly.

However, this problem formulation does not take into account the connection between two subsequent steps $t_1$ and $t_2$. That is why, the authors of \cite{Lubich_TI_2017} consider instead the \textit{dynamical tensor approximation} $\mathcal{Y}(t) \in \mathcal{M}_r$ determined
from the condition that for every time step - t the derivative $\dot{\mathcal{Y}}(t)$ which lies in the tangent space $\mathcal{T}_{\mathcal{Y}(t)}\mathcal{M}_r$ can be chosen as:

\begin{equation}\label{eq:main_obj}
    \begin{dcases} 
        \|\dot{\mathcal{Y}}(t) - \dot{\mathcal{X}}(t)\| \rightarrow \min_{\dot{\mathcal{Y}}(t) \in \mathcal{T}_{\mathcal{Y}(t)}\mathcal{M}_r} &  \\ 
      \mathcal{Y}(0) = \mathcal{R}(0) &  
    \end{dcases} 
\end{equation}
%By the way, for a given $\mathcal{Y}(t)$, the derivative $\dot{\mathcal{Y}}(t)$ is obtained in \eqref{eq:main_obj} by a linear projection, though onto a state-dependent vector space.

Moreover, in this work we suppose that the number of users $M$ and items $N$ also depends on time - $t$ such that:

\begin{equation}\label{eq:dyn_users_items}
    \begin{cases}
        M = M(t)\text{,  } M(t_1) \le M(t_2)\text{,  } t_1 \le t_2 &  \\ 
        N = N(t)\text{,  } N(t_1) \le N(t_2)\text{,  } t_1 \le t_2  &  
    \end{cases} 
\end{equation}
In this regard, we can consider four distinct cases of user and item interactions:
\begin{enumerate}
    \item Interactions with new users
    \item Interactions with new items
    \item Interactions with known users and known items
    \item Interactions with new users and new items
\end{enumerate}

The overall purpose of the study is to solve the problem \eqref{eq:main_obj}, \eqref{eq:dyn_users_items} that will allow to develop a tensor-based recommender system that can update user and items embeddings effectively. The developed model is supposed to work with data chunks rather than all the data up to the moment and should be able to add embeddings for new users and items.

\section{Dynamic Collaborative Filtering}

\begin{figure}[t]
    \centering
    \includegraphics[scale=0.29]{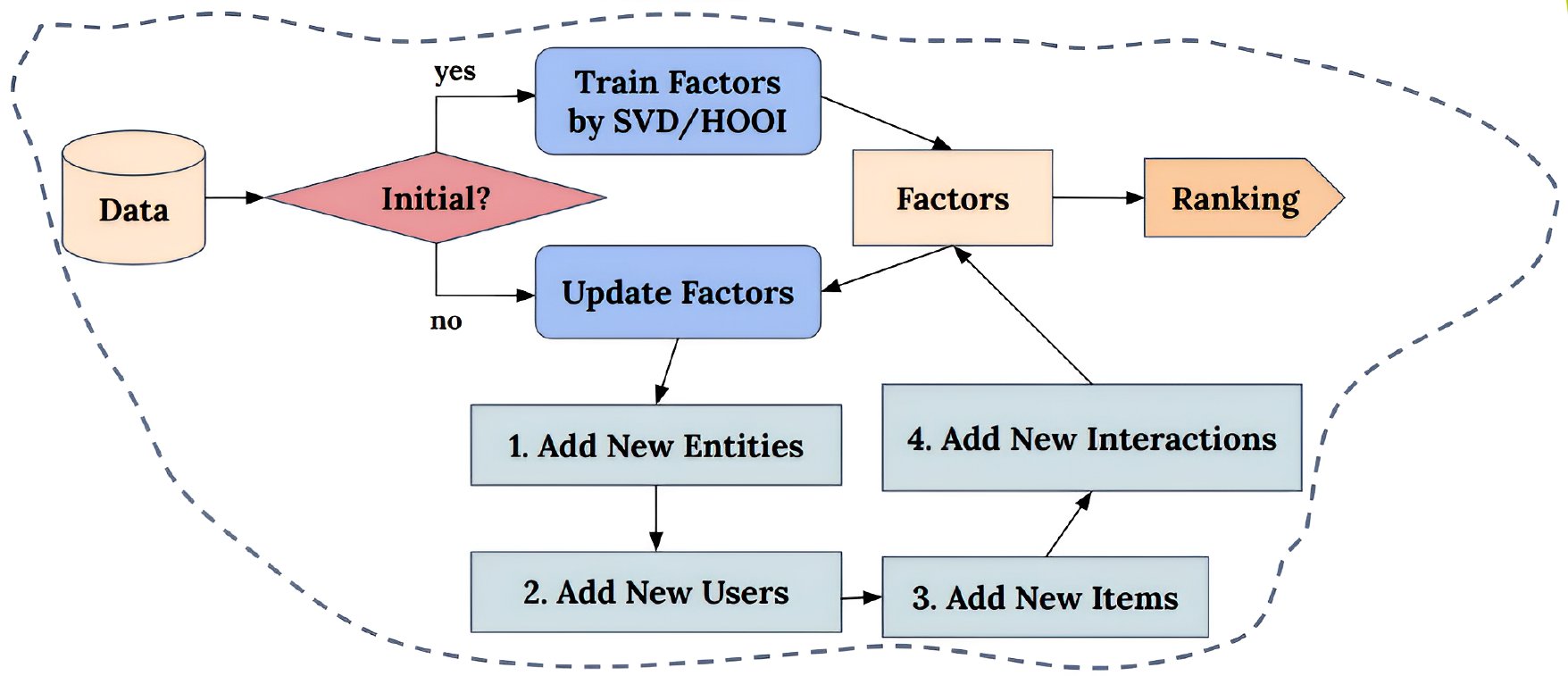}
    \caption{General integrator-based recommender system framework. \textit{New entities} mean new users that interacted only with new items. \textit{New users} represent a set of new users that interacted with known items. \textit{New items} contain a set of new items that were consumed by the known users. \textit{New interactions} represent interactions of known entities.}
    \label{fig:integrator_framework}
\end{figure} 

Let us introduce the methodology to solve the problem~\eqref{eq:main_obj}, \eqref{eq:dyn_users_items} presented on Figure~\ref{fig:integrator_framework}.
Firstly, we have some initial user-item interactions data. Using it we can find user and item embeddings using SVD or HOOI depending on dimensionality - which are model parameters. With these parameters the model can generate recommendations for a user. 
Secondly, the recommender system is dynamic because in a day or a week the new data can be generated (e.g. new users register on a platform, people create more content, people consume new content). That is why at some point the recommender model should be trained with the incoming data. In the framework of matrix and tensor decompositions one would have to either retrain the model from scratch or use time window to create a new model. In the framework of integrators we suggest to use the four step procedure depicted on Figure~\ref{fig:integrator_framework}.

The next section describes technical details needed for implementation of this framework in matrix and tensor cases: \textit{how to train initial factors}; \textit{how to add new entities, users, items}; \textit{how to add new interactions}; \textit{how to do the ranking of items for a particular user}.

\subsection{Matrix Case}\label{sec:matrix_case}
First of all, in two-dimensional case one would find the initial user $U$ and item $V$ embeddings using an SVD of incoming data matrix~\cite{Cremonesi_Topn_2010}.

\begin{algorithm}[t]
\caption{PSIRec: Update known user/item factors}
\label{alg:psi}
\begin{algorithmic}[1]
\Require \\
    - $\Delta A$ - new data collected since the last time-step for known users/items\\
    - $U_0$, $S_0$, $V_0^\top$ - last state of the model at time $t-1$ 
\Ensure $U_1$, $S_1$, $V_1^\top$ - new state of the model at time $t$
%\Function{psi}{$\Delta A$, $U_0$, $S_0$, $V_0^\top$}
    \State $U_1$, $S'$ = QR($U_0S_0 + \Delta A V_0$)
    \State $L = V_0(S' - U_1^\top\Delta A V_0)^\top + \Delta A^\top U_1$
    \State $V_1$, $S_1^\top$ = QR($L$) 
    
    %\Return $U_1, S_1, V_1^\top$
    %\EndFunction
\end{algorithmic}
\end{algorithm}

Secondly, in this research we present several techniques to incrementally add embeddings for new users and/or new items (step 2-3 on Figure~\ref{fig:integrator_framework}).
\textit{The first proposition} is to use zero embeddings as initialization for further update by PSI - Algorithm~\ref{alg:psi}. Zero initialization of new factors does conflict with the orthogonality condition for the factor matrices - $U$, $V$.
\textit{The second proposition} is to use random initialization for new user/item embeddings. As the addition of new rows - $\hat{U}$, $\hat{V}$ to matrices - $U$, $V$ would break the orthogonality condition one have to apply QR decomposition to recover the structure (example for user factors):
\begin{equation}\label{eq:qr_new_entities}
    \begin{dcases}
    \begin{bmatrix}
            U \\
            \hat{U} 
        \end{bmatrix} = QR \\
    U \leftarrow Q \\ 
    S \leftarrow RS
    \end{dcases}
\end{equation}
These new embeddings are further updated by Algorithm~\ref{alg:psi} on the forth step.
\textit{The third proposition} is to apply the idea
of incremental SVD developed by~\cite{brand_ISVD_2002}. You can find the implementation details in Algorithm~\ref{alg:vectors}.
To add new users that interacted with known items one has to swap $U$ and $V$ in Algorithm~\ref{alg:vectors} and swap resulting factors accordingly.

Another key thing to remember is that there might be a situation when a new user that is not represented in a recommender system interacted with a new item.
In this case, user and item embeddings can be updated using an SVD of a block-diagonal matrix represented in Algorithm~\ref{alg:matrix} which is the first step in update factors procedure on Figure~\ref{fig:integrator_framework}. 

\begin{algorithm}[t]
\caption{PSIRec: Add new users and new items}
\label{alg:matrix}
\begin{algorithmic}[1]
\Require \\
    - $\Delta A$ - data representing new users and new items interactions\\
    - $U_0$, $S_0$, $V_0^\top$ - state of the model 
\Ensure $U_3$, $S_3$, $V_3^\top$ - updated state of the model
%\Function{update\_matrix\_2d}{$\Delta A$, $U_0$, $S_0$, $V_0^\top$}
    \State $U_1$, $S_1$, $V_1^\top$ = SVD ($\Delta A$)
    
    \State{$U_2, S_2, V_2^\top$ = SVD
    $\left(
        \begin{bmatrix}
            S_0 & 0\\
            0 & S_1
        \end{bmatrix}
    \right)$}
    
    \State{$U_3 \leftarrow \begin{bmatrix}
            U_0 & 0\\
            0 & I
        \end{bmatrix}U_2$}

    \State{$S_3 \leftarrow S_2$}
    \State{$V_3 \leftarrow \begin{bmatrix}
            V_0 & 0\\
            0 & I
        \end{bmatrix}V_2$}

    %\Return $U_3, S_3, V_3^\top$
    %\EndFunction
\end{algorithmic}
\end{algorithm}

The final step in update factors procedure is to add new interactions. This step uses Projector-Splitting Integrator (PSI)~\cite{Lubich_PSI_2013} technique. PSI offers an
incremental scheme that takes only the newly added data to directly update the model, which eliminates the need to
recompute the entire model from scratch. See Algorithm~\ref{alg:psi} for
implementation details. 

Speaking of model inference, it is the same as for a PureSVD model~\cite{Cremonesi_Topn_2010}. Hence, given some user preferences binary vector $p \in \mathbb{R}^{M(t)}$, the list of top-n recommendations can be generated as:
\begin{equation}
    \text{toprec}(p, n) = \argmax^{n} VV^\top p,
\end{equation}
where $V$ - item embeddings matrix, $\text{toprec}(p, n)$ - returns $n$ item indices of maximum relevance score.
The presented model is called as \textbf{PSIRec}.

\subsection{Tensor Case}\label{sec:tensor_case}
According to GA-SATF model from~\cite{frolov_tensorbased_2022} to mimic
the positional attention mechanism and make the model be able to learn from sequential data we would use the following data in this section: 
\begin{equation}\label{eq:att_data_tensor}
    \mathcal{X}(t) \equiv \mathcal{X}(t) \times_{3} A^\top,
\end{equation}
where $A \in \mathbb{R}^{L\times L}$ - attention matrix such that: 
\begin{equation}
    A = \begin{bmatrix}a_1&0& \dots&0\\a_2&a_1&\dots&0\\\vdots&\vdots&\ddots&\vdots\\a_L&\dots&a_2&a_1\end{bmatrix}
\end{equation}
$a_L = L^{-f}$, $f \ge 0$ - scaling hyper-parameter.

To begin with, to get the initial embeddings for users and items we apply HOOI~\cite{Phan_randomized_TD_2021} to three dimensional sequential data tensor~\eqref{eq:att_data_tensor}.

Secondly, to add new users and/or new items (step 2-3 on Figure~\ref{fig:integrator_framework}) to a model one can apply different strategies which are the same as for a matrix case~\ref{sec:matrix_case}. However, the third proposition should be extended on tensor case which is presented in Algorithm~\ref{alg:vectors_3d}.

Thirdly, in order to add new users that interacted with new items and vise versa we have to extend Algorithm~\ref{alg:matrix} on tensor case. The incoming data can be naturally represented as a block tensor and the update procedure is depicted in Algorithm~\ref{alg:matrix_3d}.

\begin{algorithm}[t]
\caption{TIRec: Add new users and new items}
\label{alg:matrix_3d}
\begin{algorithmic}[1]
\Require \\
    - $\Delta\mathcal{A}$ - data representing new users and new items ordered interactions\\
    - $U_1^0$, $U_2^0$, $U_3^0$, $\mathcal{C}^0$ - state of the model 
\Ensure $U_1^1, U_2^1, U_3^1, \mathcal{C}^1$ - updated state of the model
%\Function{update\_matrix\_3d}{$\Delta\mathcal{A}$, $U_1^0$, $U_2^0$, $U_3^0$, $\mathcal{C}^0$}
    \State $U_1$, $V_1$, $\mathcal{C}_1$ = Tucker2 ($\Delta\mathcal{A}\times_{3}(U_3^0)^\top$)
    
    \State{$\hat{U}, \hat{V}, \hat{W}, \mathcal{C}^1$ = HOSVD
    $\left(
        \begin{bmatrix}
            \mathcal{C}^0 & 0\\
            0 & \mathcal{C}_1
        \end{bmatrix}
    \right)$}
    
    \State{$U_1^1 \leftarrow \begin{bmatrix}
            U_1^0 & 0\\
            0 & U_1
        \end{bmatrix}\hat{U}$}

    \State{$U_3^1 \leftarrow U_3^0\hat{W}$}
    \State{$U_2^1 \leftarrow \begin{bmatrix}
            U_2^0 & 0\\
            0 & V_1
        \end{bmatrix}\hat{V}$}

    %\Return $U_1^1, U_2^1, U_3^1, \mathcal{C}^1$
   % \EndFunction
\end{algorithmic}
\end{algorithm}

The forth step is to add new interactions in update factors procedure on Figure~\ref{fig:integrator_framework}. This step uses Tucker Integrator (TI)~\cite{Lubich_TI_2017} to dynamically update user, item, positional embeddings. See Algorithm~\ref{alg:tucker_integrator} for implementation details.

Speaking of model inference, it is the same as for a GA-SATF model~\cite{frolov_tensorbased_2022}. Hence, given some user preferences binary matrix $P \in \mathbb{R}^{M(t) \times L}$, the list of top-n recommendations can be generated as:
\begin{equation}
    \text{toprec}(P, n) = \argmax^{n} VV^\top PSAW\hat{w}_{L}
\end{equation}
where $V$ - item embeddings matrix, $W$ - position (order) embedding matrix, $A$ - attention matrix, $\text{toprec}(P, n)$ - returns $n$ item indices of maximum relevance score, $\hat{w} = [A^{-\top}W]_{L}$.
$S = [\delta_{k, k'+1}]\in \mathbb{R}^{L \times L}$ - lower shift matrix that decreases positions of all observed items in $P$ by one. Note that if length of a user sequence is exactly $L$, the first item in the sequence gets discarded, which satisfies the length-$L$ requirement for user histories in tensor construction.
The presented model that uses the third proposition (see~\ref{sec:matrix_case}) is called as \textbf{TIRec}. The model that uses the first proposition is called \textbf{TIRecA}.

\begin{algorithm}[t]
\caption{TIRec: Update known user/item factors}
\label{alg:tucker_integrator}
\begin{algorithmic}[1]
\Require \\
    - $\Delta\mathcal{A}$ - new data collected since the last time-step for known users/items\\
    - $U_1^0$, $U_2^0$, $U_3^0$, $\mathcal{C}^0$ - state of the model 
\Ensure $U_1^1, U_2^1, U_3^1, \mathcal{C}^1$ - updated state of the model
%\Function{tucker\_integrator}{$\Delta\mathcal{A}$, $U_1^0$, $U_2^0$, $U_3^0$, $\mathcal{C}^0$}
    \For{$i \in (1, 2, 3)$}
        \State $Q_i^0$, $(S_i^0)^\top = \text{QR}\left((\mathcal{C}^0_{[i]})^\top\right)$ 
        \State $(V_i^0)^T = \left((Q_i^0)^\top_{[i]}\bigtimes_{k=1}^{i-1}(U_k^1)^\top \bigtimes_{l=i+1}^{3}(U_l^0)^\top\right)_{[i]}$
        \State $K_i^1 = U_i^0 S_i^0 + \Delta\mathcal{A}_{[i]} V_i^0$
        \State $U_i^1$, $S_i^1 = \text{QR}\left(K_i^1\right)$
        \State $S_i^0 = S_i^1$
        \State $S_i^1 \leftarrow S_i^0 - (U_i^1)^\top \Delta\mathcal{A}_{[i]} V_i^0$
        \State $\mathcal{C}^0 = (S_i^1(Q_i^0)^\top)_{[i]}$
    \EndFor
    \State $\mathcal{C}^1 = \mathcal{C}^0 + \Delta\mathcal{A} \bigtimes_{i=1}^{3}(U_i^1)^\top$
    
    %\Return $U_1^1, U_2^1, U_3^1, \mathcal{C}^1$
    %\EndFunction
\end{algorithmic}
\end{algorithm}

\section{Experiments}\label{sec:experimental_methodology}
In this section we describe the general evaluation setup, datasets and preprocessing steps for performing experiments.

\subsection{Data}\label{sec:data}
In this work, we use the following publicly available datasets in recommender systems field: Movielens-20M (ML-20M), 
Amazon Beauty (AMZ-B), Amazon Toys and Games
(AMZ-G), and Steam \cite{McAuley_SASRec_2018}. We follow the same data initial preparation procedure as in~\cite{frolov_tensorbased_2022}. Namely, we leave no less than five interactions per each user and each item for all datasets except Movielens-20M. The explicit values of ratings are transformed into an implicit binary signal indicating the presence of a rating. As in the Steam data some users assigned more than one review to the same items, such duplicates were removed.
In Movielens-20M data we use 20\% of the last interactions.
The resulting statistics for the datasets are provided in
Table~\ref{table:data_stats}.

\subsection{Metrics}\label{sec:metrics}
In order to compare the prediction quality of different methods we would use the following metrics: HR@n, MRR@n, WJI(u, v)@n defined below: 
\begin{equation}\label{eq:hr}
    \text{HR}@\text{n} = \dfrac{\lvert U_{\text{hit}}^{n} \rvert}{\lvert U_{\text{all}} \rvert}
\end{equation}
where $\lvert U_{\text{hit}}^{n} \rvert$ is the number of users for which the correct answer is included in the top-n recommendation list, $\lvert U_{\text{all}} \rvert$ is the total number of users in the test set.

\begin{equation}\label{eq:mrr}
    \text{MRR}@\text{n} = \dfrac{1}{\lvert U_{\text{all} \rvert}} \sum_{u = 1}^{\lvert U_{\text{all}} \rvert} \dfrac{\text{relevance}(u)}{\text{rank}(u)}
\end{equation}
$\text{relevance}(u)$ is 1 if a test item is included in the top-n recommendation list else 0, $\text{rank}(u)$ is a position of a test item in the top-n recommendation list, $\lvert U_{\text{all}} \rvert$ is the total number of users in the test set.

In order to measure the stability of recommendations we would use the method from~\cite{Olaleke_Dynamic_Modeling_2021}. The top-n similarity between two recommendations lists is measured as the Weighted Jaccard Index (WJI) of their "bag-of-items" (BOI) representations $u$, $v$: 
\begin{equation}\label{eq:wji}
    \text{WJI}(u, v)@\text{n} = \dfrac{\sum_{j=1}^N \min(w_j(u), w_j(v))}{\sum_{j=1}^N \max(w_j(u), w_j(v))}
\end{equation}
where $u = (u_i)_{i=1}^{N}$ and $u_i$ denotes the position of item $i$ in the list of recommendations; $N$ - the number of items; $w_i(u) = \dfrac{1}{u_i}$ and $w_i(u) = 0$ if $u_i > n$.

\subsection{Evaluation}\label{sec:evaluations}
\begin{figure}[t]
    \centering
    \includegraphics[scale=0.34]{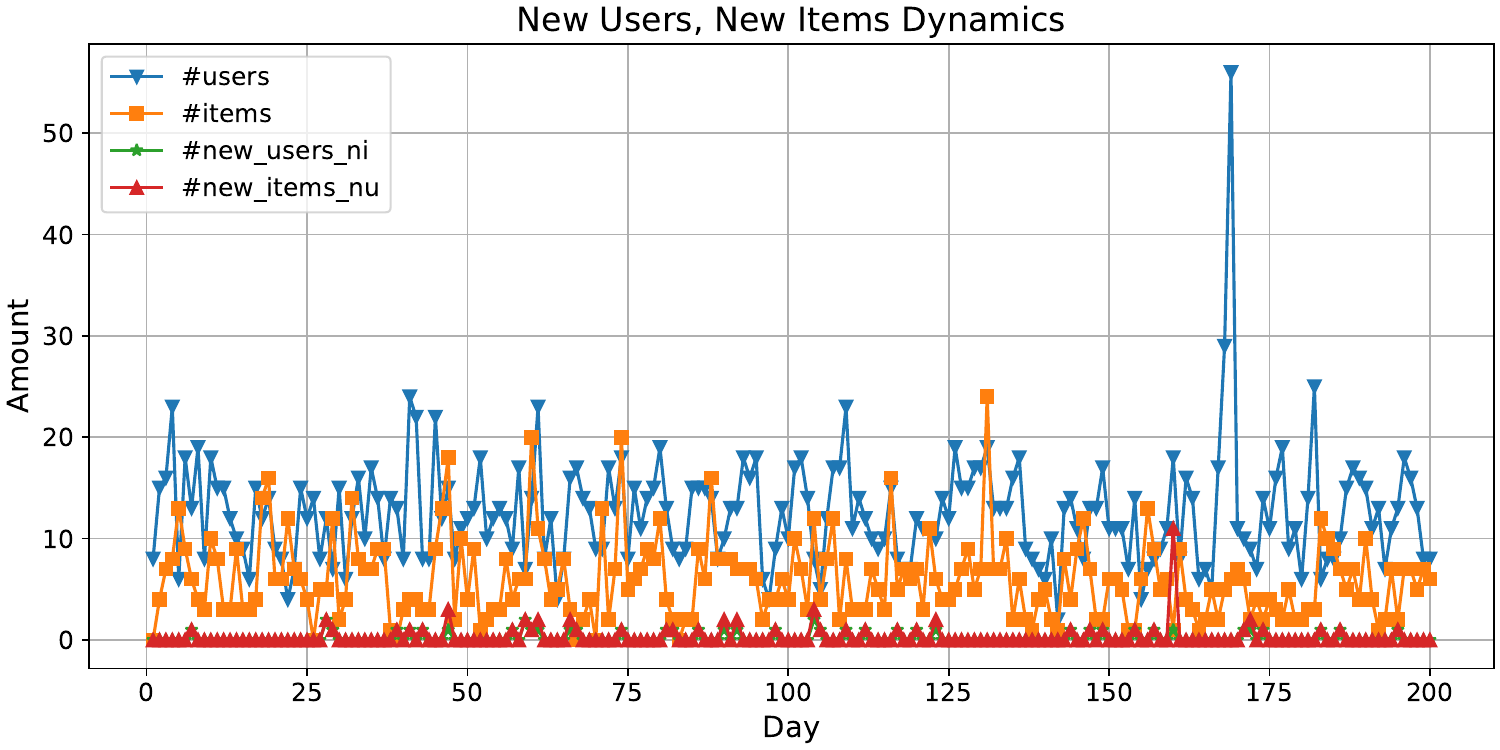}
    \caption{\textbf{ML-20M}. How many users/items are processed by a recommender system. \#user - the number of new users; \#items - the number of new items; \#new\_users\_ni - the number of new users that interacted with new items; \#new\_items\_nu - the number of new items that were consumed only by new users.}
    \label{fig:ml_20m_dynamics_ui}
\end{figure}

\begin{figure}[t]
    \centering
    \includegraphics[scale=0.34]{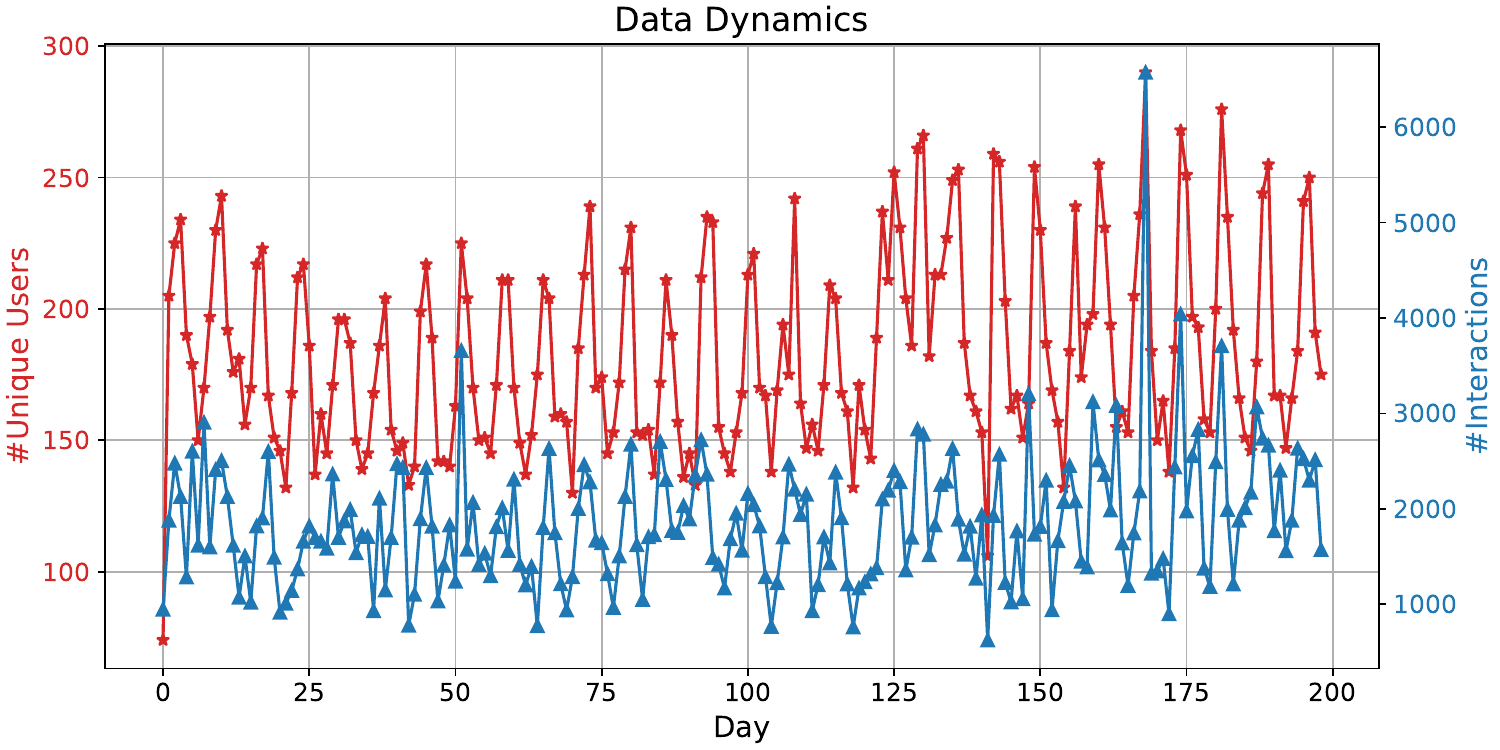}
    \caption{\textbf{ML-20M}. Dynamical nature of data in recommender systems field. }
    \label{fig:ml_20m_data_dynamics}
\end{figure}

As in a general Machine Learning field, we preprocess the data and split it into train, validation and test subsets. Validation is used for hyper-parameters optimization, whereas test is used for final evaluation. In our settings test data is divided into chunks - chunk$_1$, chunk$_2$, \dots, etc sequentially day by day. You can find data dynamic on Figures \ref{fig:ml_20m_dynamics_ui},  \ref{fig:ml_20m_data_dynamics} for ML-20M dataset.

The first step is to merge the validation data back into the train data when the optimal hyper-parameters are found. Then, we can retrain the model using train data and optimal hyper-parameters. After that, we can evaluate the model on chunk$_1$. To do that, we have to find a set of users that are present in the train data and chunk$_1$. The first items these users interacted with in chunk$_1$ are the target we are going to compare against the recommended lists of items using metrics defined above in~\ref{sec:metrics}. Then we can retrain the model using all the data up to the moment - train and chunk$_1$ or update the model parameters using only chunk$_1$ which depends on the model. The same procedure is applied several times to different chunks sequentially. In the end one can get the dynamical behavior of the model based on measured metrics.

In order to evaluate a recommender system's stability we find a set of target users beforehand. They must be present in train data and in many chunks. Top-50 users are chosen for that purpose. To measure stability we compare ranking lists from two subsequent time steps for target users and take an average by their number.

For Amazon Beauty and Amazon Toys and Games the following settings were used: 70\% of initial data goes into train; 0.01\% out of train is kept for validation; 100 chunks (days) were used for dynamical evaluation; maximum length of a user sequential history - $K = 20$.
Experimental setup for Movielens-20M data is: 40\% of initial data goes into train; 0.01\% out of train is kept for validation; 200 chunks (days) were used for dynamical evaluation; maximum length of a user sequential history - $K = 20$.
For experiments with the final Steam dataset we used the settings:
80\% of initial data goes into train; 0.01\% out of train is kept for validation; 50 chunks (days) were used for dynamical evaluation; maximum length of a user sequential history - $K = 8$;

\subsection{Baselines}\label{sec:baselines}
The main idea of this work is to develop faster alternatives to SVD-based models and models based on Tucker Decomposition. 
The first baseline is \textbf{PureSVD} model (SVD)~\cite{Cremonesi_Topn_2010}. It is a strong baseline in recommender systems field. In the settings of this work, this model is retrained with all the data up to date.
The second baseline is \textbf{TDRec}. This model applies general Tucker Decomposition to incoming data tensor and finds factors and core out of it. This model is also retrained from scratch every time step using all the gathered data.
The final baseline is \textbf{TDRecReinit}. This model applies Tucker Decomposition to incoming data as well. But on the next time step it uses the factors from the previous iteration that allows for faster and more robust retraining.

\subsection{Hyper-Parameters Grid Search}\label{sec:hyperparams}
In this work we compare different models that can be categorized into two types: matrix-based and tensor-based. 

Matrix-based models are \textbf{PureSVD} and \textbf{PSIRec} and they have 1 hyper-parameter: rank r. For rank values r, their range is (10, \dots , 300) with step size 10 for all datasets but Steam. For Steam data, rank values range is (10, \dots , 400) with step size 10.

Tensor-based models are \textbf{TDRec}, \textbf{TDRecReinit}, \textbf{TIRec}, \textbf{TIRecA}. These models have 2 main hyper-parameters: rank \textbf{r} = $(r_1, r_2, r_3)$ and scaling parameter $f$ (see~\ref{sec:tensor_case}). For rank values $r_1$, $r_2$, their range is (32, 64, 100, 128, 256), $r_3$ range is (5, 10), $f \in (0, 2, 4)$ for all datasets but Steam. For Steam data, rank values $r_1 \in (64, 128, 256)$, $r_2 \in (64, 100)$, $r_3 \in (2, 4)$ and scaling factor $f \in (0, 2)$.  

In order to find optimal hyper-parameters we fix parameters configuration, then train the model using training data and measure prediction quality using validation data. After that we can find the best parameters configuration based on metric results. 
The source code to
reproduce our work is openly published online~\footnote{\url{https://github.com/AlbMLpy/DynamicCF}}.

\section{Results and Discussion}
This section presents experimental comparison of various models in terms of computational time, prediction quality and stability of recommendations. Apart from that we compare different regimes to add embeddings for new users and/or items in the proposed frameworks. We present graphical results only for ML-20M dataset due to space restrictions. The average results for all other data can be find in the following tables: Table~\ref{table:ev_results} shows comparison of proposed models with the baselines~\ref{sec:baselines}; Table~\ref{table:tirec_configs} presents the metrics results for different options to add new embeddings to a model based on Tucker Integrator; Table~\ref{table:psirec_configs} presents the metrics results for different options to add new embeddings to a model based on PSI.

\subsection{Computational Time Results}\label{sec:compute_results}

\begin{figure}[t]
    \centering
    \includegraphics[scale=0.34]{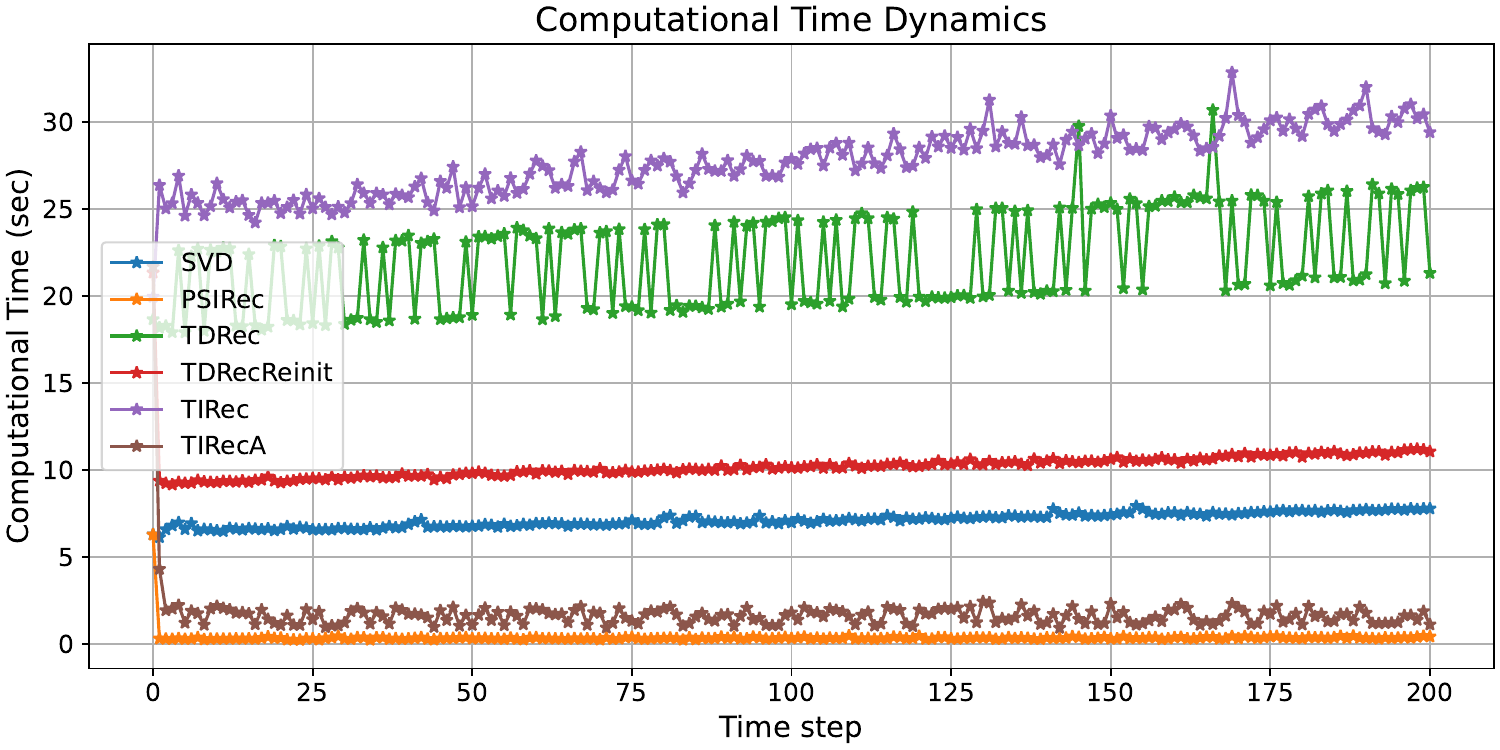}
    \caption{\textbf{ML-20M}. Computational time dynamics}
    \label{fig:ml_20m_calc_time}
\end{figure}

Let us start from the comparison of the training time per iteration across different datasets and models.
As we can conclude from the Figure~\ref{fig:ml_20m_calc_time} the slowest model in terms of computational time is TIRec. The main problem of TIRec is the addition of new users or new items Algorithm~\ref{alg:vectors_3d}. Being mathematically correct, it is a bottleneck of the whole method which prevents scalability. One have to calculate an SVD of a wide dense matrix (see line 5 in Algorithm~\ref{alg:vectors_3d}). The second to worst model is TDRec. It is expected as the model is retrained from scratch every time step with all the data up to the moment. It is worth mentioning that the fluctuations in the green curve could be a result of different number of iterations in the inner cycle of TDRec training procedure (see HOOI~\cite{Kolda_TD_2009}). TDRecReinit baseline needs much less iterations to converge then TDRec because of proper initialization and it allows to accelerate the model. The results for TIRecA model confirm that removing Algorithm~\ref{alg:vectors_3d} from TIRec allows to get increase in training speed. This model is about 25 times faster then TIRec, 9 times faster then TDRecReinit, 20 times faster then TDRec and even 3 times faster then SVD (see Table~\ref{table:ev_results}). The best model according to the plot is PSIRec. The results for the other three datasets, AMZ-B and AMZ-G, Steam are similar to the ML-20M data and can be found in an accumulated way in Table~\ref{table:ev_results}.

\subsection{Prediction Quality Results}\label{sec:pred_qual_results}

\begin{figure}[t]
    \centering
    \includegraphics[scale=0.34]{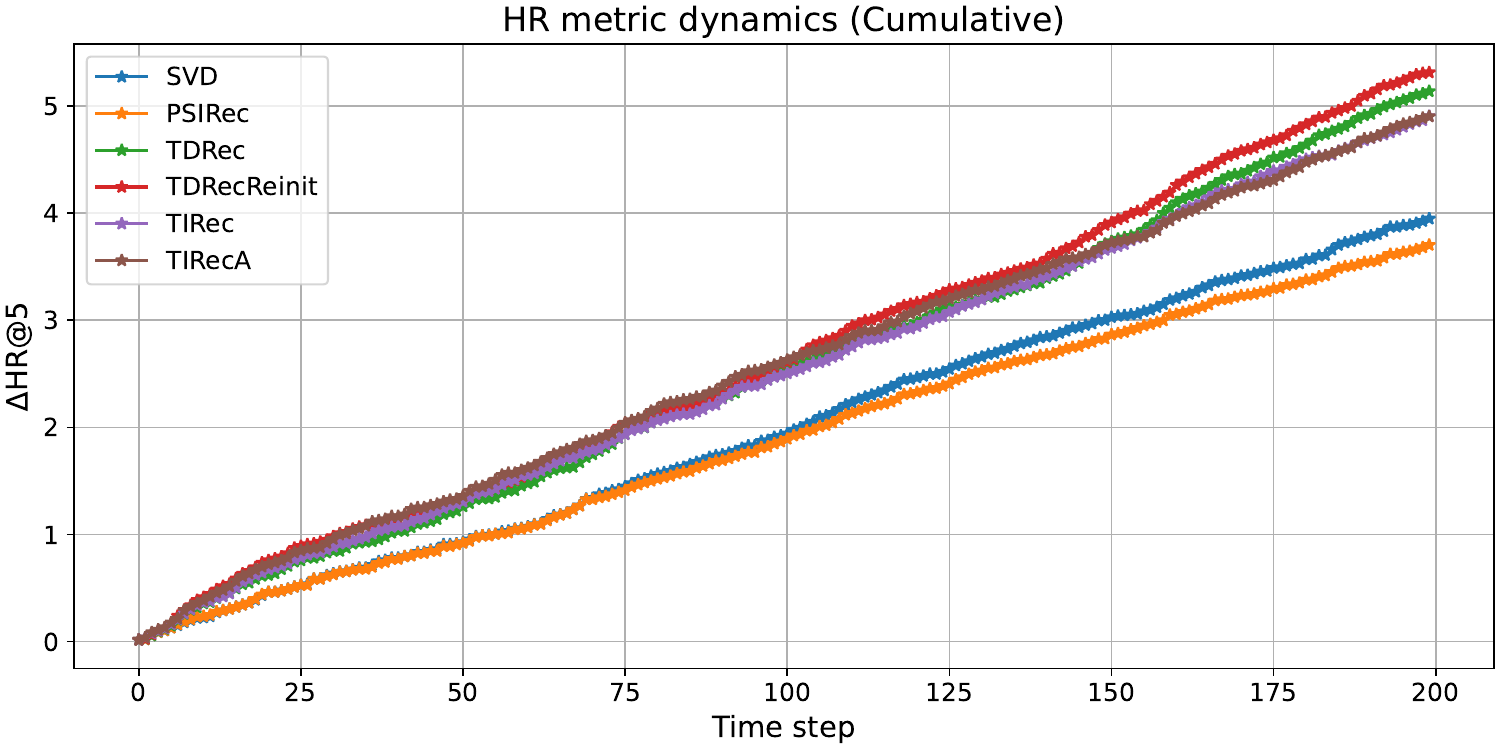}
    \caption{\textbf{ML-20M}. Dynamics of HR metric (Cumulative)}
    \label{fig:ml_20m_cum_hr}
\end{figure}

Let us further analyse the results of HR metric dynamics for ML-20M dataset on Figure~\ref{fig:ml_20m_cum_hr}. As one can see the sequential information is important in this particular dataset as the difference between matrix and tensor models is significant. In its turn, PSIRec model results are very similar to SVD baseline. The results for tensor-based models are also close to each other with best model - TDRecReinit. This graph shows the behavior of integrator-based models that we wanted to achieve - the prediction quality is close to the baselines.
According to Table~\ref{table:ev_results}, one can conclude that the sequential information is not that important for Steam and AMZ-G datasets as PSIRec and SVD show better HR@5 metric results than tensor counterparts. The best tensor model on these data is TDRecReinit. For AMZ-B dataset one can see a major difference in prediction quality for tensor methods: 0.0013 for TIRecA model against 0.020 in TDRecReinit. This model behavior can be possibly explained by a significant change in users' behavior which breaks the assumptions (too large step size - h) under the Tucker Integrator~\cite{Lubich_TI_2017}.

\subsection{Stability Results}\label{sec:stab_results}

\begin{figure}[t]
    \centering
    \includegraphics[scale=0.34]{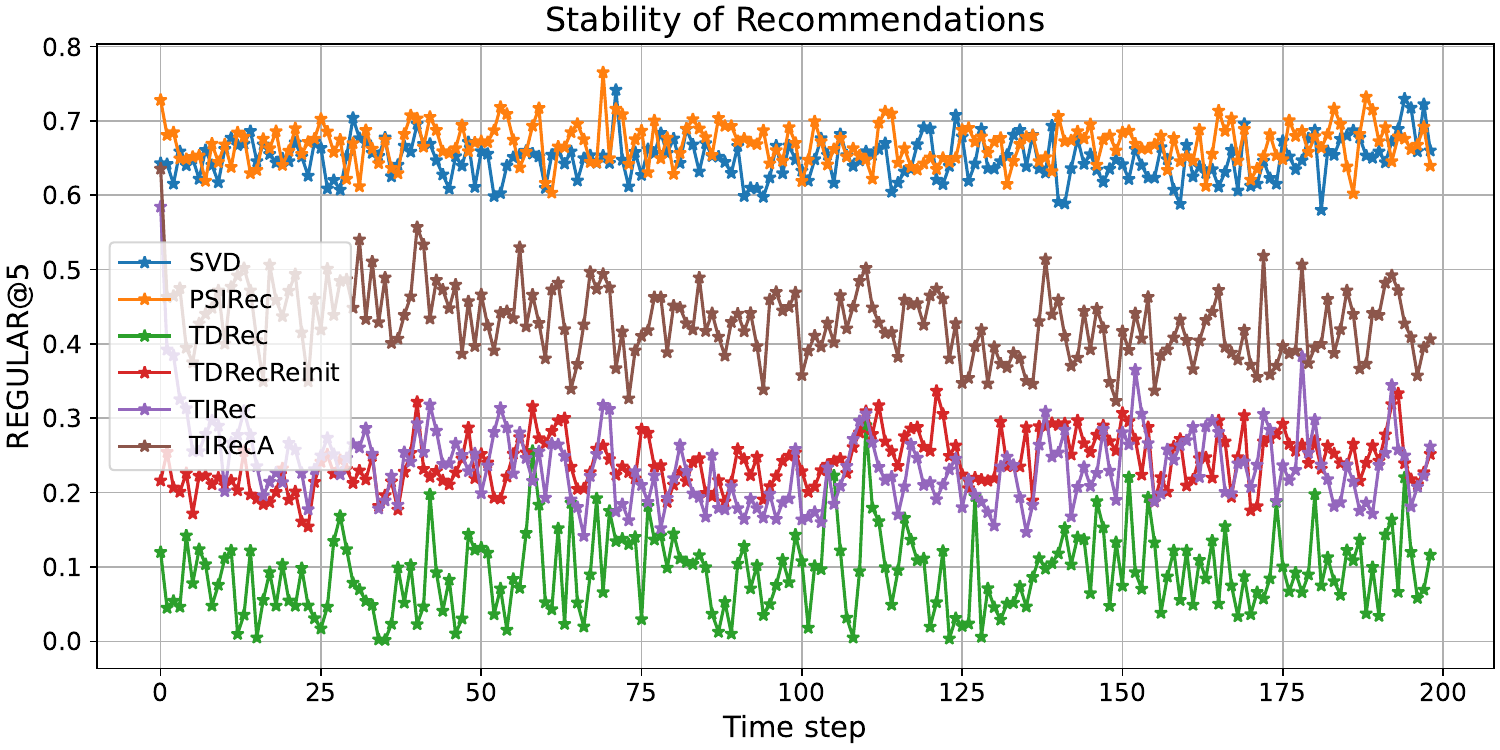}
    \caption{\textbf{ML-20M}. Stability of recommendations}
    \label{fig:ml_20m_stability}
\end{figure}
This section presents the findings on the stability of different recommender models in dynamical settings.
Consider the example of the ML-20M dataset illustrated in Figure~\ref{fig:ml_20m_stability}. The WJI@5 value indicates the stability of the model, where a larger value implies less variation in recommendations between subsequent time steps. Matrix models exhibit the best stability performance, with an average value of 0.668 for PSIRec and 0.649 for SVD (refer to Table~\ref{table:ev_results}). TIRecA performs better than the other tensor models, surpassing them by approximately twofold. TDRec is the least stable model, with instances where the WJI metric reaches 0, indicating a completely different recommendation list compared to the previous time step. That is not a stable recommendation. The result on stability for the other datasets can be found in Table~\ref{table:ev_results}). We can conclude that the most stable matrix-based model is PSIRec and tensor-based one is TIRecA most of the time.

\subsection{Update Methods Results}\label{sec:upd_propositions}
This section presents the results of different initialization strategies for new user/item factors in integrator-based framework (see~\ref{sec:matrix_case}).

\begin{figure}[t]
    \centering
    \includegraphics[scale=0.34]{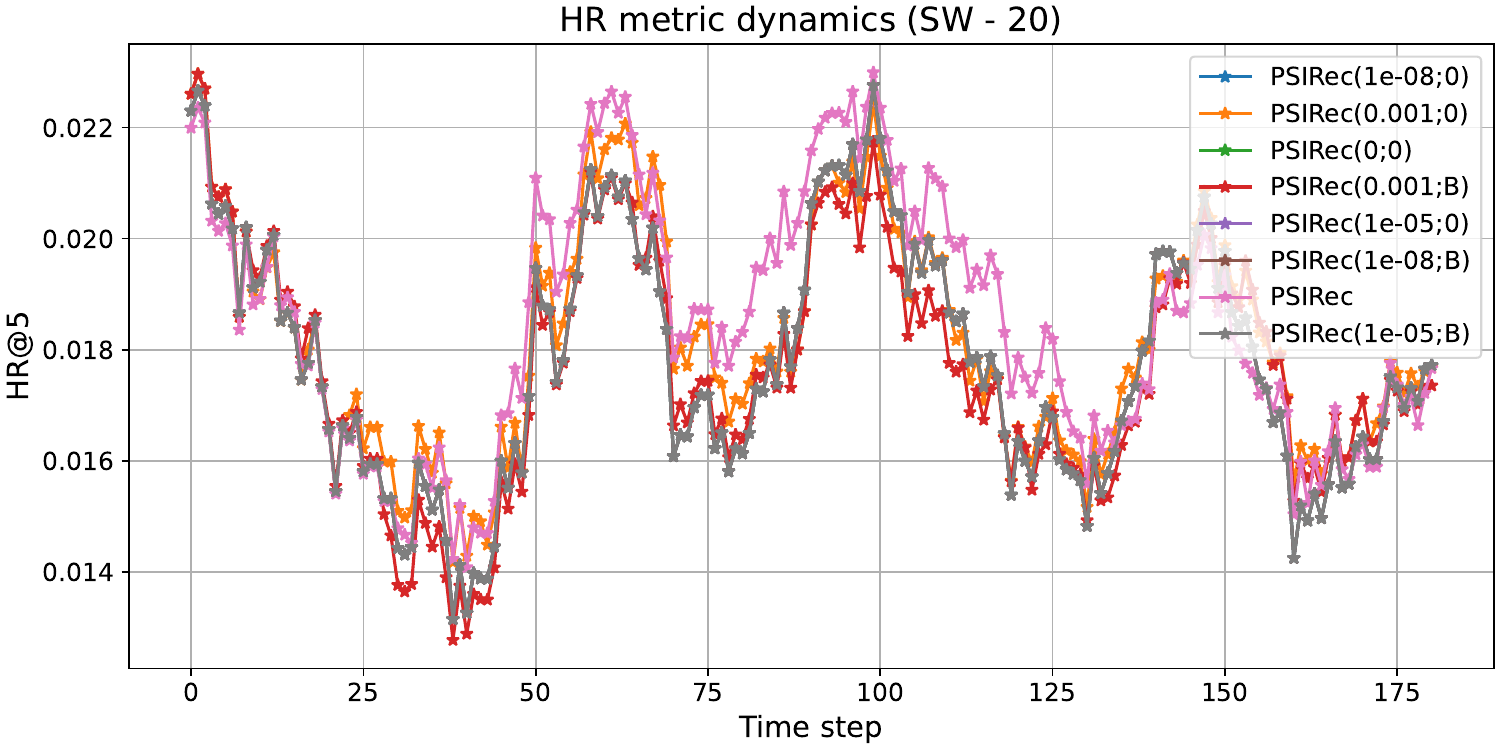}
    \caption{\textbf{ML-20M}. Comparison of different initialization strategies for PSIRec model. PSIRec($\sigma$;$Y$) depicts different model configurations, where $\sigma$ taken from $\mathcal{N}(0,\,\sigma^{2})$ initialization of factors; $Y = 0$ - zero initialization in the first step of Figure~\ref{fig:integrator_framework}, $Y = \text{B}$ - use Algorithm~\ref{alg:matrix}. The results are represented using a sliding window with step 1 and window length 20.}
    \label{fig:ml_20m_psirec_compare}
\end{figure}

Let's start with PSIRec model on Figure~\ref{fig:ml_20m_psirec_compare}. As we can see random initialization does not improve the prediction quality of a model. The best one is pure PSIRec using Algorithms~\ref{alg:vectors}, \ref{alg:matrix} to add new embeddings to a system. The same behavior holds true for all the datasets (see Table~\ref{table:psirec_configs}). It should be noted that, the fastest models are PSIRec and PSIRec(0;0) as they do not use extra QR decomposition. That is why, these methods are preferred as fast and accurate.

\begin{figure}[t]
    \centering
    \includegraphics[scale=0.34]{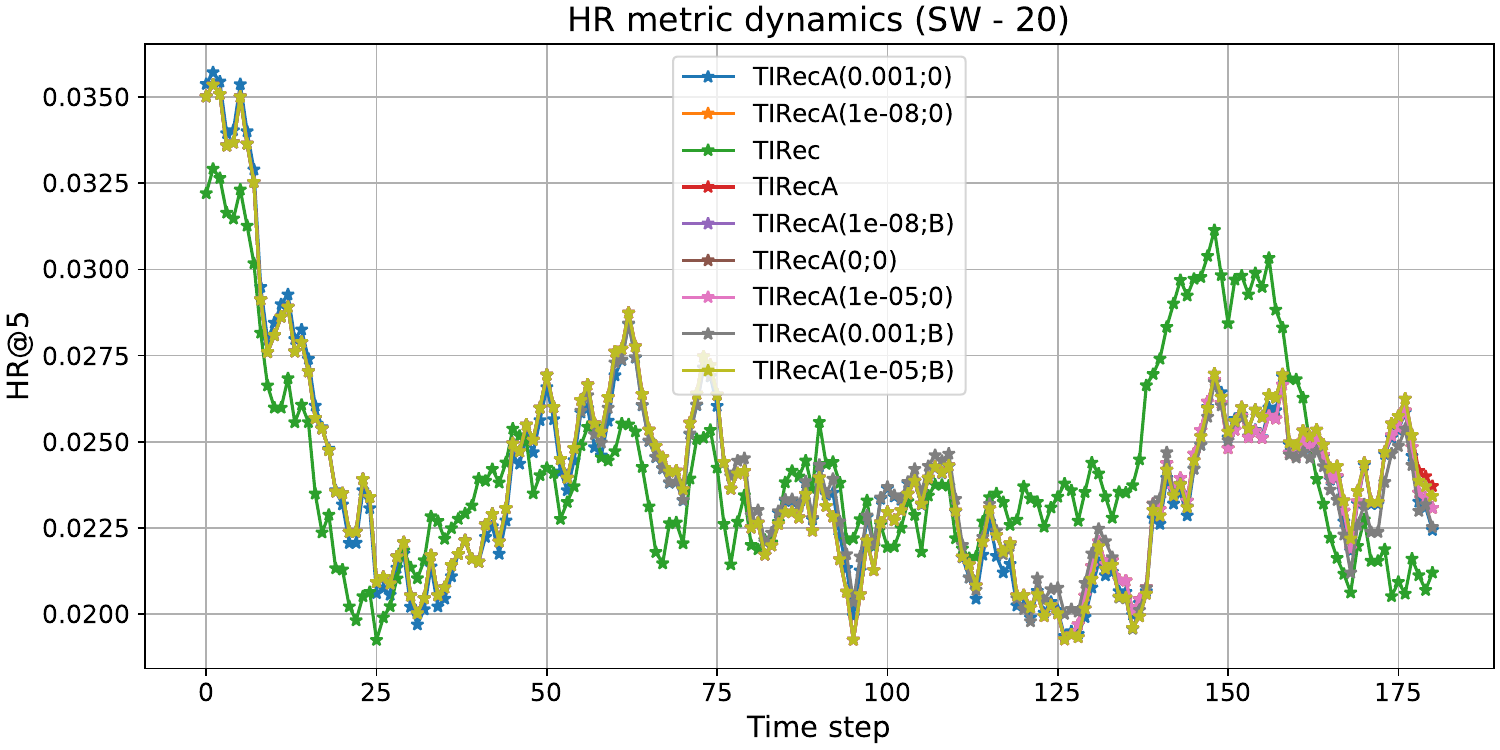}
    \caption{\textbf{ML-20M}. Comparison of different initialization strategies for TIRec model. TIRecA($\sigma$;$Y$) depicts different model configurations, where $\sigma$ taken from $\mathcal{N}(0,\,\sigma^{2})$ initialization of factors; $Y = 0$ - zero initialization in the first step of Figure~\ref{fig:integrator_framework}, $Y = \text{B}$ - use Algorithm~\ref{alg:matrix_3d}. The results are represented using a sliding window with step 1 and window length 20.}
    \label{fig:ml_20m_tireca_compare}
\end{figure}

Now let's consider TIRec model results depicted on Figure~\ref{fig:ml_20m_tireca_compare} and Table~\ref{table:tirec_configs}. The best model in terms of HR metric quality is TIRec on ML-20M data. However, on Amazon data the models based on random initialization show more accurate results - TIRecA(0.001;B) and TIRecA(1e-05;0). Speaking of training time, the best configuration is TIRecA(0;0) as it uses pure Tucker Integrator to update the user/item factors.

\section{Conclusion}
The main purpose of this work was to design and implement a novel sequential recommender system in the framework of tensor data representation that would be able to update its parameters effectively using incoming data and would save appropriate prediction quality. This model was developed and it is called - \textbf{TIRecA}. Comprehensive set of experimental results confirms that this model can be 10-20 times faster than the general models based of Tucker decomposition. Speaking of metric results, the model showed comparable performance to all the baselines under different settings. However, it should be noted that the model could degrade in the long run and in order to restore the performance one have to retrain the model from scratch periodically. 

Aforementioned results provide a good starting point for discussion and further research in the area of sequential recommender systems. One possible future research topic could be combination of the proposed TIRecA model and model based on Neural architecture. The idea is to use the training speed of tensor model and combine it with prediction quality of contemporary neural models. Another interesting topic would be an adaptation of TIRecA model to a four-dimensional tensor data used in~\cite{frolov_tensorbased_2022} in order to accelerate the LA-SATF model training.

\section*{Acknowledgment}
This work was supported by the Russian Science Foundation under Grant 22-21-00911.
%The preferred spelling of the word ``acknowledgment'' in America is without an ``e'' after the ``g''. Avoid the stilted expression ``one of us (R. B. G.) thanks $\ldots$''. Instead, try ``R. B. G. thanks$\ldots$''. Put sponsor acknowledgments in the unnumbered footnote on the first page.

\bibliographystyle{IEEEtran}
\bibliography{IEEEabrv, main}

\appendix

%%% TABLES %%%

\begin{table}[h!]
\caption{Datasets statistics after initial preprocessing}
\begin{center}
\begin{tabular}{||l||c|c|c|c|c||}
\hline
Dataset &  n\_users &  n\_items &  n\_actions &  density\% \\
\hline
 ML-20M &    31670 &    26349 &         4000052 &      0.48 \\
  AMZ-B &    22363 &    12101 &          198502 &      0.07 \\
  AMZ-G &    19412 &    11924 &          167597 &      0.07 \\
  Steam &   281206 &    11961 &         3484694 &      0.10 \\
\hline
\end{tabular}
\label{table:data_stats}
\end{center}
\end{table}

\begin{table}
\caption{Results of evaluation. All metrics are averaged by days and computed for top-n recommendations with n = 5}
\label{table:ev_results}
\begin{tabular}{||l|l|c|c|c|c||}
\hline
 &  & ML-20M & AMZ-B & AMZ-G & Steam \\
\hline
Time & PSIRec & \textbf{0.376} & \textbf{0.722} & \textbf{0.591} & \textbf{0.657} \\
 & SVD & 7.129 & 4.420 & 1.513 & \underline{5.413} \\
 & TDRec & 22.064 & 10.028 & 12.550 & 130.491 \\
 & TDRecReinit & 10.223 & 5.288 & 5.303 & 52.141 \\
 & TIRec & 27.681 & 39.237 & 31.858 & - \\
 & TIRecA & \underline{1.729} & \underline{1.253} & \underline{1.284} & 17.410 \\
\hline
HR & PSIRec & 0.019 & 0.013 & \underline{0.016} & \textbf{0.022} \\
 & SVD & 0.020 & \underline{0.015} & \textbf{0.017} & \textbf{0.022} \\
 & TDRec & \underline{0.026} & \textbf{0.020} & 0.012 & 0.015 \\
 & TDRecReinit & \textbf{0.027} & \textbf{0.020} & 0.013 & \underline{0.020} \\
 & TIRec & 0.025 & 0.012 & 0.011 & - \\
 & TIRecA & 0.025 & 0.013 & 0.010 & 0.009 \\
\hline
MRR & PSIRec & 0.008 & 0.007 & \underline{0.008} & \textbf{0.011} \\
 & SVD & 0.009 & 0.008 & \textbf{0.009} & \textbf{0.011} \\
 & TDRec & \textbf{0.013} & \underline{0.010} & 0.006 & 0.007 \\
 & TDRecReinit & \textbf{0.013} & \textbf{0.011} & 0.006 & \underline{0.010} \\
 & TIRec & \underline{0.012} & 0.006 & 0.006 & - \\
 & TIRecA & \textbf{0.013} & 0.007 & 0.005 & 0.004 \\
\hline
WJI & PSIRec & \textbf{0.668} & \textbf{0.673} & \textbf{0.663} & \textbf{0.683} \\
 & SVD & \underline{0.649} & \underline{0.641} & \underline{0.624} & \underline{0.672} \\
 & TDRec & 0.092 & 0.030 & 0.029 & 0.111 \\
 & TDRecReinit & 0.241 & 0.231 & 0.167 & 0.488 \\
 & TIRec & 0.236 & 0.214 & 0.229 & - \\
 & TIRecA & 0.429 & 0.224 & 0.210 & 0.539 \\
\hline
\end{tabular}
\end{table}

\begin{table}
\caption{Comparison of different initialization strategies for PSIRec model. All metrics are averaged and computed for top-n recommendations with n = 5. PSIRec($\sigma$;$Y$) depicts different model configurations, where $\sigma$ taken from $\mathcal{N}(0,\,\sigma^{2})$ initialization of factors; $Y = 0$ - zero initialization in the first step of Figure~\ref{fig:integrator_framework}, $Y = \text{B}$ - use Algorithm~\ref{alg:matrix}.}
\label{table:psirec_configs}
\begin{tabular}{||l|l|c|c|c|c||}
\hline
 &  & ML-20M & AMZ-B & AMZ-G & Steam \\
\hline
Time & PSIRec & 0.376 & 0.722 & 0.591 & 0.657 \\
 & PSIRec(0.001;0) & 0.677 & 1.397 & 1.164 & 0.690 \\
 & PSIRec(0.001;B) & 0.676 & 1.423 & 1.206 & 0.707 \\
 & PSIRec(0;0) & 0.326 & 0.731 & 0.619 & 0.417` \\
 & PSIRec(1e-05;0) & 0.666 & 1.389 & 1.185 & 0.690 \\
 & PSIRec(1e-05;B) & 0.636 & 1.404 & 1.194 & 0.700 \\
 & PSIRec(1e-08;0) & 0.631 & 1.393 & 1.178 & 0.674 \\
 & PSIRec(1e-08;B) & 0.690 & 1.409 & 1.181 & 0.686 \\
\hline
HR & PSIRec & 0.019 & 0.013 & 0.016 & 0.022 \\
 & PSIRec(0.001;0) & 0.018 & 0.012 & 0.015 & 0.022 \\
 & PSIRec(0.001;B) & 0.018 & 0.012 & 0.015 & 0.022 \\
 & PSIRec(0;0) & 0.018 & 0.013 & 0.016 & 0.022 \\
 & PSIRec(1e-05;0) & 0.018 & 0.013 & 0.016 & 0.022 \\
 & PSIRec(1e-05;B) & 0.018 & 0.013 & 0.016 & 0.022 \\
 & PSIRec(1e-08;0) & 0.018 & 0.013 & 0.016 & 0.022 \\
 & PSIRec(1e-08;B) & 0.018 & 0.013 & 0.016 & 0.022 \\
\hline
MRR & PSIRec & 0.008 & 0.007 & 0.008 & 0.011 \\
 & PSIRec(0.001;0) & 0.009 & 0.007 & 0.008 & 0.010 \\
 & PSIRec(0.001;B) & 0.008 & 0.006 & 0.008 & 0.011 \\
 & PSIRec(0;0) & 0.008 & 0.007 & 0.008 & 0.010 \\
 & PSIRec(1e-05;0) & 0.008 & 0.007 & 0.008 & 0.011 \\
 & PSIRec(1e-05;B) & 0.009 & 0.007 & 0.008 & 0.011 \\
 & PSIRec(1e-08;0) & 0.008 & 0.007 & 0.008 & 0.010 \\
 & PSIRec(1e-08;B) & 0.009 & 0.007 & 0.008 & 0.010 \\
\hline
WJI & PSIRec & 0.668 & 0.673 & 0.663 & 0.683 \\
 & PSIRec(0.001;0) & 0.667 & 0.668 & 0.670 & 0.679 \\
 & PSIRec(0.001;B) & 0.673 & 0.674 & 0.665 & 0.676 \\
 & PSIRec(0;0) & 0.666 & 0.672 & 0.665 & 0.678 \\
 & PSIRec(1e-05;0) & 0.669 & 0.673 & 0.660 & 0.678 \\
 & PSIRec(1e-05;B) & 0.669 & 0.677 & 0.667 & 0.678 \\
 & PSIRec(1e-08;0) & 0.664 & 0.674 & 0.666 & 0.678 \\
 & PSIRec(1e-08;B) & 0.672 & 0.673 & 0.666 & 0.680 \\
\hline
\end{tabular}
\end{table}

\begin{table}
\caption{Comparison of different initialization strategies for TIRec model. All metrics are averaged and computed for top-n recommendations with n = 5. TIRecA($\sigma$;$Y$) depicts different model configurations, where $\sigma$ taken from $\mathcal{N}(0,\,\sigma^{2})$ initialization of factors; $Y = 0$ - zero initialization in the first step of Figure~\ref{fig:integrator_framework}, $Y = \text{B}$ - use Algorithm~\ref{alg:matrix_3d}.}
\label{table:tirec_configs}
\begin{tabular}{|l|l|c|c|c|c||}
\hline
 &  & ML-20M & AMZ-B & AMZ-G & Steam \\
\hline
Time & TIRec & 27.681 & 39.237 & 31.858 & - \\
 & TIRecA & 1.729 & 1.253 & 1.284 & 17.410 \\
 & TIRecA(0.001;0) & 1.677 & 1.799 & 1.360 & 27.723 \\
 & TIRecA(0.001;B) & 2.079 & 2.171 & 2.246 & 27.847 \\
 & TIRecA(0;0) & 1.397 & 1.300 & 1.044 & 16.855 \\
 & TIRecA(1e-05;0) & 1.673 & 1.816 & 1.350 & 27.665 \\
 & TIRecA(1e-05;B) & 2.091 & 2.188 & 2.261 & 27.837 \\
 & TIRecA(1e-08;0) & 1.675 & 1.810 & 1.345 & 27.305 \\
 & TIRecA(1e-08;B) & 2.086 & 2.187 & 2.274 & 27.911 \\
\hline
HR & TIRec & 0.025 & 0.012 & 0.011 & - \\
 & TIRecA & 0.025 & 0.013 & 0.010 & 0.009 \\
 & TIRecA(0.001;0) & 0.024 & 0.014 & 0.010 & 0.009 \\
 & TIRecA(0.001;B) & 0.024 & 0.015 & 0.011 & 0.009 \\
 & TIRecA(0;0) & 0.024 & 0.014 & 0.011 & 0.009 \\
 & TIRecA(1e-05;0) & 0.024 & 0.014 & 0.012 & 0.009 \\
 & TIRecA(1e-05;B) & 0.024 & 0.014 & 0.011 & 0.009 \\
 & TIRecA(1e-08;0) & 0.024 & 0.014 & 0.011 & 0.009 \\
 & TIRecA(1e-08;B) & 0.024 & 0.014 & 0.011 & 0.009 \\
\hline
MRR & TIRec & 0.012 & 0.006 & 0.006 & - \\
 & TIRecA & 0.013 & 0.007 & 0.005 & 0.004 \\
 & TIRecA(0.001;0) & 0.012 & 0.007 & 0.005 & 0.004 \\
 & TIRecA(0.001;B) & 0.012 & 0.008 & 0.005 & 0.004 \\
 & TIRecA(0;0) & 0.012 & 0.007 & 0.005 & 0.004 \\
 & TIRecA(1e-05;0) & 0.012 & 0.007 & 0.006 & 0.004 \\
 & TIRecA(1e-05;B) & 0.012 & 0.008 & 0.006 & 0.004 \\
 & TIRecA(1e-08;0) & 0.012 & 0.007 & 0.005 & 0.004 \\
 & TIRecA(1e-08;B) & 0.012 & 0.007 & 0.006 & 0.004 \\
\hline
WJI & TIRec & 0.236 & 0.214 & 0.229 & - \\
 & TIRecA & 0.429 & 0.224 & 0.210 & 0.539 \\
 & TIRecA(0.001;0) & 0.424 & 0.223 & 0.210 & 0.536 \\
 & TIRecA(0.001;B) & 0.425 & 0.219 & 0.214 & 0.542 \\
 & TIRecA(0;0) & 0.427 & 0.226 & 0.206 & 0.539 \\
 & TIRecA(1e-05;0) & 0.427 & 0.222 & 0.208 & 0.547 \\
 & TIRecA(1e-05;B) & 0.425 & 0.220 & 0.210 & 0.534 \\
 & TIRecA(1e-08;0) & 0.427 & 0.225 & 0.206 & 0.545 \\
 & TIRecA(1e-08;B) & 0.429 & 0.219 & 0.209 & 0.542 \\
\hline
\end{tabular}
\end{table}

%%% TABLES %%%

%%% ALGORITHMS %%%

\begin{algorithm}[t]
\caption{PSIRec: Add new users or new items}
\label{alg:vectors}
\begin{algorithmic}[1]
\Require \\
    - $\Delta A$ - data representing new users/items as columns\\
    - $U_0$, $S_0$, $V_0^\top$ - state of the model 
\Ensure $U_1$, $S_1$, $V_1^\top$ - updated state of the model
%\Function{update\_vectors\_2d}{$\Delta A$, $U_0$, $S_0$, $V_0^\top$}
    \State{J, K = QR($\Delta A - U_0U_0^\top \Delta A$)} 
    \State{$U', S', V'^\top$ = SVD$\left(
        \begin{bmatrix}
            S_0 & U_0^\top \Delta A\\
            0 & K
        \end{bmatrix}
    \right)$}
    
    \State $U_1 \leftarrow [U_0 J]U'$ 
    \State $S_1 \leftarrow S'$
    \State $V_1 \leftarrow \begin{bmatrix}
            V_0 & 0\\
            0 & I
        \end{bmatrix}V'$
    
    %\Return $U_1, S_1, V_1^\top$
    %\EndFunction
\end{algorithmic}
\end{algorithm}

\begin{algorithm}[t]
\caption{TIRec: Add new users or new items}
\label{alg:vectors_3d}
\begin{algorithmic}[1]
\Require \\
    - $\Delta\mathcal{A}$ - data tensor representing new users or new items interactions\\
    - $U_1^0$, $U_2^0$, $U_3^0$, $\mathcal{C}^0$ - state of the model\\
    - $m$ - target mode: 1 for new users, 2 for new items
\Ensure $U_1^1, U_2^1, U_3^1, \mathcal{C}^1$ - updated state of the model
%\Function{add\_entities}{$\Delta\mathcal{A}$, $U_1^0$, $U_2^0$, $U_3^0$, $\mathcal{C}^0$, $m$}
    \State{$\mathcal{C}_{[m]}^0 \left(U_3^0 \otimes U_2^0\right)^\top =  \Tilde{U}\Tilde{S}\Tilde{V}^\top$} 
    \State $\hat{U} = U_1^0\Tilde{U}$
    \State{$\bar{V}, \bar{S}, \bar{U}^\top = $ update\_vectors\_2d($\Delta\mathcal{A}_{[m]}^\top$, $\Tilde{V}$, $\Tilde{S}$, $\hat{U}^\top$)}
    \State $\mathcal{C}_{[m]}^1 = \bar{S} \bar{V}^\top \left(U_3^0 \otimes U_2^0\right)$
    
    %\Return $\bar{U}, U_2^0, U_3^0, \mathcal{C}^1$
    %\EndFunction
\end{algorithmic}
\end{algorithm}

%%% ALGORITHMS %%%

\end{document}